\documentclass{aa}
\input{psfig.sty}
\begin{document}

\thesaurus{11(11.03.1; 11.12.2)} 
\title{Composite luminosity function of cluster galaxies}

\author{Bianca Garilli\inst{1} \and Dario Maccagni\inst{1} \and Stefano Andreon\inst{2}}
\institute{Istituto di Fisica Cosmica G. Occhialini, via Bassini 15, 20133 Milano, 
Italy; Electronic mail: bianca,dario@ifctr.mi.cnr.it \and
Osservatorio Astronomico di Capodimonte, via Moiariello 16, 80131 Napoli, Italy;
Electronic mail: andreon@na.astro.it}

\offprints{B. Garilli}
\mail{Istituto di Fisica Cosmica G. Occhialini, via Bassini 15, 20133 Milano, 
Italy}

\date{Received / Accepted }

\maketitle

\begin{abstract}
We constructed the composite Luminosity Function (LF) of cluster galaxies in the
$g$, $r$ and $i$ bands from the photometry of a mixed (Abell and X-ray selected)
sample of the cores of 65 clusters, ranging in redshift from 0.05 to 0.25.
The composite LF has been obtained from complete samples of $\sim2200$
galaxies in the magnitude range $-23<M<-17.5$ ($-18$ in $i$). Cluster membership has
been determined on the basis of color-color plots for each cluster and the
resulting outlier counts have been checked
against field counts in the same bands. We find that the galaxy density of the environment
determines the shape of the LF, in the sense that bright galaxies are more
abundant in dense clusters.  
\end{abstract}

\keywords{galaxies, clusters --- galaxies, luminosity function}

\section{Introduction}
The study of the galaxy luminosity function (LF) in clusters has
at least two purposes: (1) to look for differences in the LF of
the different clusters, according to their different dynamical status;
(2) to compare the galaxy LF in clusters and in the field, and thus
to study the influence of the environment on the global statistical
properties of the galaxies.\\

The first cumulative cluster galaxy LF dates back
to 1976 (Schechter 1976). Later, Lugger (1983) found that the average LF of 9 clusters
was well described by a Schechter function with parameters $M_R^*=-22.74\pm0.10$ and $\alpha=-1.27\pm0.04$ in the magnitude
range $-24.5<M_R<-20$. Here and in the following we adopt
$H_0=50$ km s$^{-1}$ Mpc$^{-1}$ and
$q_0=0.5$. All absolute magnitudes have consequently been
converted to the long distance scale.\\
 
Gaidos (1997) constructed a galaxy LF from R imaging
of 20 Abell clusters and also found that it is well described by
a Schechter function with parameters $M_R^*=-22.63\pm0.11$ and
$\alpha=-1.09\pm0.08$ in the magnitude range $-24.91<M_R<-18.91$.
Clusters had redshifts in the range $0.06<z<0.25$.
Gaidos' composite galaxy cluster LF has a slope similar to the
field LF derived from the Las Campanas Redshift Survey (Lin et al. 1996),
but the value of $M^*$ is almost one magnitude brighter. To our knowledge
this composite cluster galaxy LF is the only one obtained in a red filter from CCD
imaging. 

Valotto et al. (1997) have computed the cluster galaxy luminosity
function in two cluster samples. Galaxy magnitudes have been obtained
from the Edinburgh-Durham Southern Galaxy Catalogue (Heydon-Dumbleton, Collins, \& MacGillivray 1989), and are thus $b_J$
magnitudes. All clusters lie at $z<0.07$ and the limiting absolute
magnitudes are $M=-17.5$ and $M=-18.5$. For the total sample, the best fitting
Schechter function has $M^*_{b_j}=-21.5\pm0.1$ and $\alpha=-1.4\pm0.1$. There
is marginal evidence that in poor clusters galaxies have a flatter
LF.

Finally, Lumsden et al. (1997) derive the galaxy LF in the range
$-22.5<M_{b_J}<-19.5$ from a sample of 46 clusters drawn from the Edinburgh/Durham Cluster Catalogue (Lumsden et al. 1992). Cluster redshifts vary from
0.07 to 0.16. The composite LF is derived from 22 of the richer
clusters in the sample and has Schechter best fit parameters $M^*_{b_j}=-21.66\pm0.02$ and $\alpha=-1.22\pm0.04$. Differences in the 
galaxy LF between different cluster subgroups are found to be weak.\\

These recent results suggest that galaxy cluster LFs are
steeper in the blue than in the red and that their characteristic
magnitude is brighter than in the field, by approximately one
magnitude in the red (Gaidos 1997 cluster LF with respect to Lin et al.
1996 field LF), and by approximately half a magnitude in the blue
(Lumsden et al. 1997 and Valotto et al. 1997 LFs with respect to
the ESP field LF (Zucca et al. 1997)).\\
%for which $M^*_{b_j}=-21.1$ with the
%same slope found by Lumsden et al. 1997).\\

In this paper, we make use of the multicolor photometric catalog of Abell and {\it Einstein} Medium Sensitivity Survey (EMSS) clusters
of galaxies described in Garilli et al. (1996) to derive the composite
galaxy LF in the $g$, $r$ and $i$ passbands. We must stress that
the available cluster imaging is restricted to areas of size close to the
cluster cores. The
paper is organized as follows: in section 2 we describe the 
cluster subsample, summarize the photometric technique and discuss
the background subtraction; in 
section 3 we illustrate the method used to construct the composite
LFs; in section 4 we present the multicolor LFs obtained for the
total sample and for the different cluster
subgroups in which the sample can be divided; finally, in section 5
we discuss our results in the light of the recent LF determinations 
in clusters and in the field.\\

\section{The Data}
\subsection{The Cluster Sample}
The cluster sample used in this work was presented in Tables 1
and 2 of Garilli et al. (1996). In the present work, we excluded 2
clusters (A175 and A410) because their color-color diagrams are anomalous
and spectroscopy of a limited number of galaxies in those fields
(Bottini et al., in preparation) points to a high background contamination.
Three fields, respectively in A1785, A272 and A439, were also excluded 
from the present analysis because their images are shallower
than the average (cfr. Table 2 in Garilli et al. 1996).\\

Finally, galaxies are extracted from 65 clusters, 44 of which are
Abell clusters, while the remaining 21 are X-ray selected clusters
from the EMSS catalog. The cluster redshift range is $0.05<z<0.25$.
The average area covered by the CCD images of each cluster 
has a radius of $\sim350$ kpc, but with important variations from
cluster to cluster, ranging from $\sim90$ to $\sim650$ kpc.\\   

\subsection{Galaxy Photometry}
The original photometric catalog from Garilli et al. (1996) was
not produced in a fully automatic way. This has some drawbacks when
a precise computation of the completeness limit is required. 
Therefore, we have run SExtractor (Bertin \& Arnouts 1996) on
the original images, and have compared the results with the original 
catalogues: the two catalogues are virtually undistinguishable,
but for the very faintest objects.
Magnitudes in the $g$, $r$ and $i$ bands have been computed
within 10 kpc radii. The choice of a metric aperture magnitude assures the
correct computation and comparison of galaxy fluxes and colors in clusters at
different redshifts.
To allow for comparison with other works (see section 5), isophotal corrected total magnitudes in the $r$ filter have also
been computed. Given the spread in $z$ of our cluster sample, k-correction must be applied
to get consistent absolute magnitudes. K-corrections depend on spectral type,
which, at the zero order, can be assimilated to morphological type.
Reliable morphological classifications cannot be derived from our images, 
but for the brighter galaxies. A possible approach is to assume a morphological
composition and apply statistical k-corrections. In our case (small central areas
of $\sim350$ kpc radius), the morphological composition is highly skewed
towards early type galaxies ($70\%$ of E+S0 galaxies, Dressler et al. 1997). Moreover,
the peak of the redshift distribution is at $z\sim0.15$, where differences in 
the k-corrections between ellipticals and spirals are at most 0.1 mag in the
$g$ band. Thus, we chose to apply the ellipticals k-corrections (taken from
Frei \& Gunn 1994) to all galaxies.\\

\subsection{Background Subtraction}
In the most recent works, background subtraction is performed by counting galaxies 
in annuli
around the cluster positions and statistically subtracting the field
contribution to the cluster galaxy counts. Alternatively, when the detector
field of view is not large enough, flanking fields are obtained from where to
infer the local background. Our data do not allow us to follow any of these
procedures. However, we have data in three different filters and thus we
can exploit colors to remove from each cluster photometric catalog the galaxies
with colors not matching the expected ones at the cluster redshift. The method  
(cfr. Garilli, Maccagni \& Vettolani 1991, Garilli et al. 1992) assumes that
the colors of normal galaxies can be well predicted in this
redshift range and therefore galaxies with colors different
from the expected ones are interlopers. We followed a procedure which
is best illustrated on the basis of Figure 1. For each cluster, we 
plot all galaxies brighter than $m_{lim}$ (as defined in section
3.1) in the $g-r$, $r-i$ plane. We define and draw the straight line
(line $a$ in Figure 1) with angular coefficient defined by the 
k-corrected colors of 
ellipticals
and spirals (Frei \&
Gunn 1994), offset to match the colors of the three brightest galaxies in the cluster
field (assumed to be ellipticals) in order
to take into account the possible color shift with respect to the
Virgo color-magnitude relation (Garilli et al. 1996). In the same way, we can compute
the expected $g-r$ color of ellipticals: line $c$ is the perpendicular to line $a$
passing through this point, and represents the reddest color beyond which
we do not expect to find cluster galaxies. To take into account both dispersion
on the expected colors of ellipticals and statistical errors in our data, we
compute the distance
from line $c$ of all galaxies with
an $r-i$ color within $3\sigma$ of line $a$ and redder than line $c$.
The dispersion of these distances, combined with the expected 
dispersion in the $g-r$ colors of ellipticals, assumed to be 0.05 mag, 
summed to the expected
colors of ellipticals, defines line $c'$: all galaxies redder
than this value are rejected. As pointed out in 2.2, we do not expect many 
spirals in our fields, not to mention irregulars. On the other hand, we can have
some contamination from foreground field galaxies. Depending on their
redshift, these galaxies will be found in the bluer part of the diagram.
A way to get rid of most of these objects without artificially depleting
our clusters of all the spirals they might have, is to compute the minimum
expected $\Delta(g-r)$ between spirals and ellipticals at the cluster
redshift (Frei \& Gunn 1994), and draw line $b$ perpendicularly to line $a$
at that point. All galaxies bluer than line $b$ have a high probability
of being interlopers and are therefore rejected.  
Finally, we compute the robust
average distance of galaxies from line $a$, combine it with the
intrinsic color dispersion and determine lines $d$ and $e$.
All galaxies lying outside the horizontal strip defined by lines 
$d$ and $e$ are rejected.\\

\begin{figure}
\psfig{figure=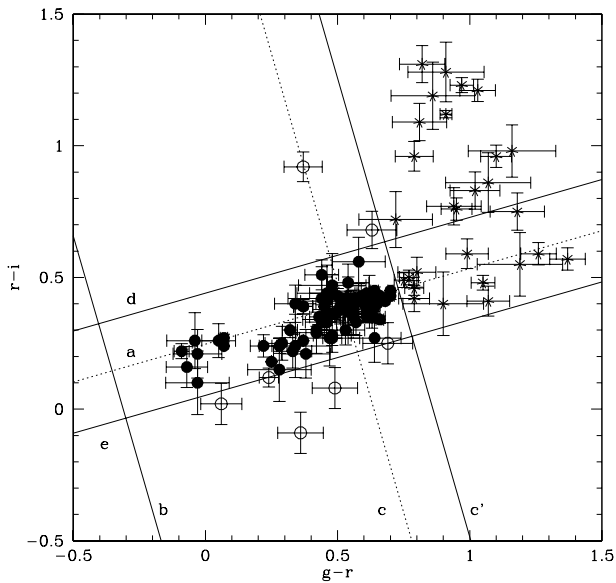,width=8cm,bbllx=20pt,bblly=165pt,bburx=570pt,bbury=700pt}
\caption{A typical cluster color-color diagram used to establish cluster membership
(see text). Filled dots: cluster member galaxies; starred symbols: background galaxies;
open circles: galaxies excluded from membership because too blue or too red in $r-i$ for 
their $g-r$ color.}
\end{figure}

We then checked the reliability of this method, applied to fields covering
the cluster cores, by comparing the results both with spectroscopic
measurements and with the field counts obtained in the same photometric system
by Neuschaefer \&
Windhorst (1995). Bottini et al. (in preparation)
measured the redshifts of 153 galaxies (m$_r\leq20$) in several of our
sample cluster fields. For 147 galaxies (96\%), the assignment based on the color-color technique
described above is spectroscopically confirmed (132 cluster members and
15 background galaxies). We erroneously included 4 galaxies (2.5\%) and we
lost 2 galaxies (1.5\%). We can therefore conclude that the adopted color-color
background subtraction method  gives quite satisfactory results for m$_r\leq20$.
The comparison with the Neuschaefer \&
Windhorst (1995) field counts is less straightforward because of the varying
completeness limits in our fields and consequently of the variation of the
surveyed areas as a function of the magnitude. Figure 2 shows the total surveyed area
as a function of the $r$ completeness limit magnitude of the cluster fields.
Note that magnitude bins are not constant. Figure 3 shows the background counts
in magnitude bins of constant area,
obtained with the color-color method outlined above, compared
with the expected counts in the $r$ band on the basis of the Neuschaefer \&
Windhorst's (1995) data. According to these authors, field count fluctuations
over areas on the order concerning us are $\sim15\%$, and this is represented
in Figure 3 by the boundary of the strips around the expected value. Errors on
our data points (which have not been rebinned) are Poissonian. Inspection of 
Figure 3 shows that the background counts we obtain are fully compatible
with the field counts of Neuschaefer \&
Windhorst (1995) up to m$_r=21$. Up to m$_r=21.5$, our counts are still
compatible, albeit with a larger scatter, while beyond this magnitude the
color-color method (applied to these specific fields) probably underestimates the
background counts. More quantitively, in order to perfectly align our background
counts to the Neuschaefer \&
Windhorst's (1995) ones, we should subtract 26 more galaxies (169 have already been considered background) out of a total
of 390 in the magnitude bin $21.0<m_r\leq21.5$, 54 more galaxies (113 have already been considered background) out of a
total of 238 in the magnitude bin $21.5<m_r\leq22.0$, and 29 more galaxies (35 have already been considered background) out
of a total of 80 in the magnitude bin $22.0<m_r\leq22.5$. Note that, in the brighter
magnitude bin, the difference is on the same order as the background fluctuations.\\

\begin{figure}
\psfig{figure=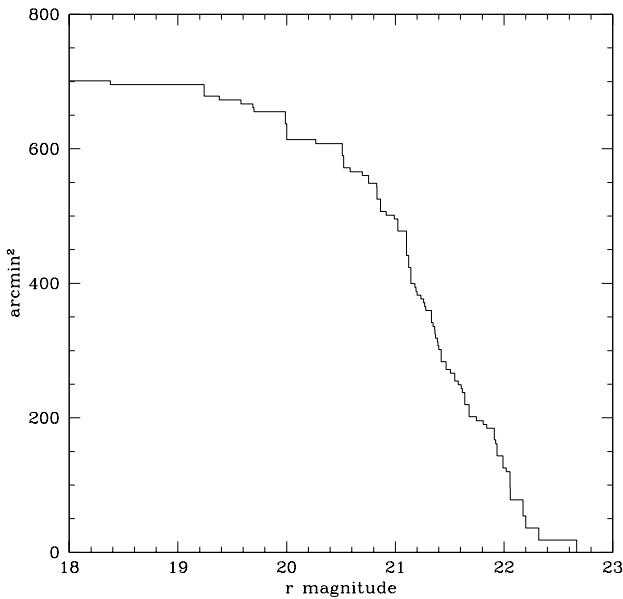,width=8cm,bbllx=35pt,bblly=160pt,bburx=570pt,bbury=700pt}
\caption{The area covered by the cluster fields as a function of the $r$ 
completeness magnitude. The area gently drops for m$_r<21$, then decreases
abruptly.}
\end{figure}

\begin{figure}
\psfig{figure=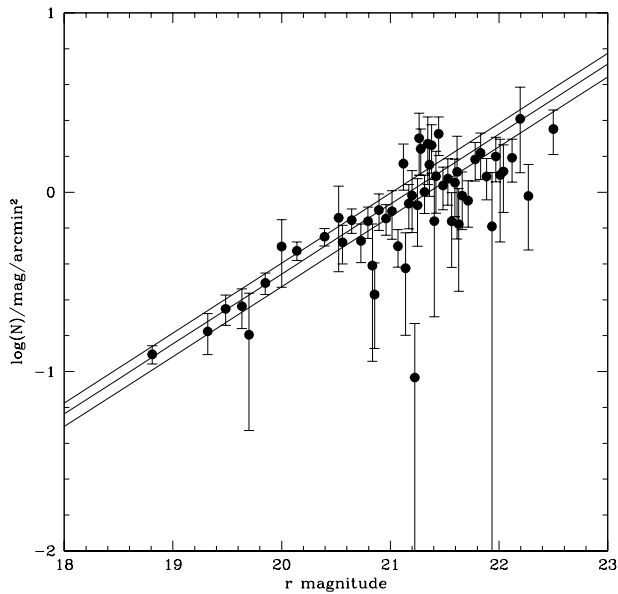,width=8cm,bbllx=35pt,bblly=160pt,bburx=570pt,bbury=700pt}
\caption{The $r$ band galaxy counts derived from non member galaxies selected
on the basis of the color-color plot of each cluster. Errors are $1\sigma$ Poissonian errors over the number of background galaxies
in equal area magnitude bins. The continuous straight lines are the 
field counts obtained by Neuschaefer \& Windhorst 1995, where we have
considered background fluctuations on the order
of 15\%.}
\end{figure}

Because of the complexity in the construction of the composite
luminosity function and of the different limiting magnitude of each
cluster catalog, it is difficult to foresee {\it a priori} how the
derived luminosity function is affected by an incorrect background
subtraction in the faint magnitude range. In order to assess the influence of
the uncertainties in the background subtraction on the LF shape, we derived
the cluster composite LF from 3 different catalogs. The first one is limited
to m$_r=21$, where both spectroscopy and comparison with field number counts confirm the reliability of our background subtraction procedure. The second one is limited
to m$_r=21.5$, where our field counts still agree within $1\sigma$ with the 
Neuschaefer \& Windhorst (1995) counts, and, finally, the third
catalog includes all the assumed cluster galaxies.
 
\section{Construction of the Composite Luminosity Function}
In order to construct the luminosity function, we need to evaluate two more quantities: the
completeness magnitude limit, $m_{lim}$, and the crowding
correction.\\ 

\subsection{Completeness}
Usually, the magnitude completeness is measured through the
detectability, as a function of the magnitude, of model galaxies which
mimic the two-dimensional
surface brightness distribution of real galaxies. In this work, 
we followed a slightly different approach: we estimated the completeness magnitude limit as the magnitude
at which we begin to loose {\it real galaxies} because they are fainter than our brightness threshold in the
detection cell.\\

The detection limit is set on the magnitude in the detection cell. The correspondence between magnitude in the detection cell and any other magnitude has a certain scatter, which depends essentially on galaxy profile. In Figure 4, line {\it a} represents 
the limit in the detection cell, while line {\it b} is the linear relation between the magnitude
within 10 kpc and the flux within the detection cell (plus and minus 
$1\sigma$, dashed lines). If the intersection between {\it a} and {\it b}
(dotted line) were taken as the completeness limit in the metric aperture, it is evident that any galaxy falling in the hatched area would not be detected even 
if brighter than the completeness limit. These "lost" galaxies become 
more numerous as
the dispersion on line {\it a} increases, and represent the low surface brightness population. This bias is minimized if, as we did, the dispersion around {\it b} is
taken into account, i.e. if we assume the continuous line as the aperture
magnitude completeness limit.\\

\begin{figure}
\psfig{figure=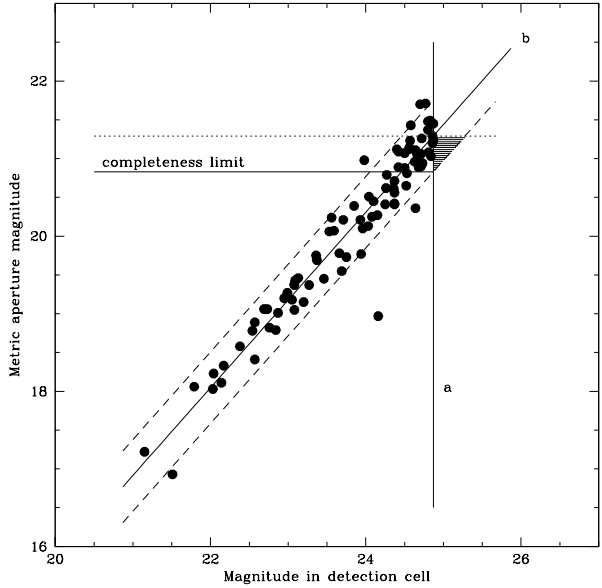,width=8cm,bbllx=35pt,bblly=165pt,bburx=570pt,bbury=700pt}
\caption{Determination of the completeness magnitude limit in a fixed aperture
given a detection limit (see text) for one of our cluster fields. The completeness limit is given by the 
intersection of line $a$ with the lower envelope of the locus of the detected galaxies.}
\end{figure}

\begin{figure}
\psfig{figure=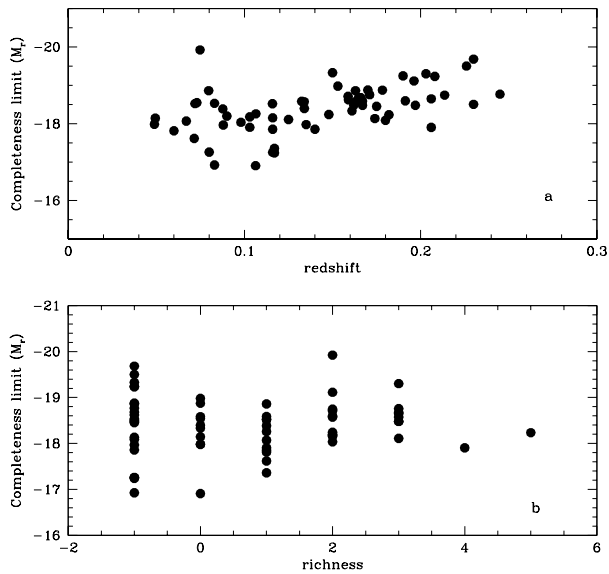,width=8cm,bbllx=20pt,bblly=160pt,bburx=570pt,bbury=700pt}
\caption{Aperture absolute magnitude ($M_r$) completeness limit as a function of redshift
(a) and as a function of richness (b) for the whole sample of clusters.}
\end{figure}

Figure 5 shows the distribution of absolute magnitudes to which our final cluster galaxy catalogs are complete. There is no trend as a function of cluster richness (b), while a slight trend with redshift is present (a). Figure 5 (a) shows that only
$\sim10\%$ of the clusters in our sample have completeness limits fainter than $M_r=-17.5$ and that none of the higher redshift clusters has a completeness limit fainter than
$M_r=-18$. For the $g$ and $i$ catalogs, completeness limits for the various clusters have the same behavior. Thus, the LFs we will derive can be considered representative of
the cluster sample for magnitudes brighter than $M_r=-17.5$, $M_g=-17.5$, and $M_i=-18$,
with the {\it caveat} that in the fainter bins galaxies are mainly drawn from clusters
at $z<0.15$.\\

\subsection{Crowding Correction}
An object, to be detected, must satisfy two conditions: its magnitude
in the detection cell must be brighter than a given threshold {\it and} the
magnitude contrast with respect to the surroundings must be above a given threshold. This second requirement is necessary to avoid multiple
detections in case of resolved structures in the objects, or of noise fluctuations in the halo of large
extended objects. When running SExtractor on our fields, this contrast was set at 5.7
mag. Therefore, objects with central surface brightness more than
5.7 mag fainter than the
one in which they are embedded are not detected at all. Since our galaxies
of lowest central surface brightness are scarcely 5.7 mag fainter than the
central surface brightness of bright cluster galaxies, this correction is expected
to be negligible for all our measurements, as we verified to be the case.\\

\subsection{The Luminosity Functions}
In most of our clusters there are too few galaxies to determine
accurately the shape of the luminosity function. On the contrary, the total
number of galaxies in all the cluster sample would allow an accurate
evaluation of the composite LF. The most straightforward way to construct a composite LF is to
add the single cluster LFs down to the brightest completeness magnitude, or to some
other appropriate limit.  Obviously, the absolute magnitude range of such an LF will be limited by the shallower of the cluster magnitude limit,
which, in our case, would lead to an inefficient use of the available data.\\

Following Colless' (1989) formulation, we constructed the composite LFs
by combining the LFs of all clusters according to:

$$N_{cj} = {{1\over m_j}\sum_{i} N_{ij}w_i}$$

\noindent 
where $N_{cj}$ is the number of galaxies in the $j$th bin of the composite
LF, $m_j$ is the number of clusters with limiting magnitude deeper than the $j$th bin, $N_{ij}$ is
the number of galaxies in the $j$th bin of the $i$th cluster, and $w_i$
is the weight of each cluster, given by the ratio of the number
of galaxies of the $i$th cluster to the number of galaxies brighter than
its magnitude limit in all clusters with fainter magnitude limits.\\

Our way of constructing the composite LFs differs from Colless' (1989) only 
in the way the weight of each contributing cluster is
computed. In order to make use of all our data base, we weigh 
clusters on the number of galaxies in an adaptive magnitude
range in order to cope with the varying cluster richness and 
surveyed areas in our sample.\\

The formal error of the composite LF is computed according to:

$$\sigma_{N_{cj}} = {{1\over m_j}\sqrt{\sum_{i} N_{ij}w_i{^2}}}$$

\noindent
The final $r$ band complete cluster galaxy catalog respectively contain 2265 
galaxies, 2154 of which are brighter than m$_r=21.5$ and 1971 are brighter
than m$_r=21$. Completeness limits have been evaluated independently in each filter,
and since all galaxies brighter than the detection limit in one
filter have also been detected in the other filters, our catalogs
are complete to the respective magnitude limit independently of the
galaxy colors. At the limiting magnitude, the signal to noise
ratio is still $\sim15$. The brightest
galaxy in each cluster has been removed from the catalogs.\\

\section{Results}
We first derived the composite LFs in each filter from each of the apparent magnitude 
limited galaxy catalogs (see section 2.3). Magnitude limits in the $g$ and $i$ 
bands have been
estimated following the same criteria used for the $r$ band, i.e. compatibility
levels with the field counts of Neuschaefer \& Windhorst (1995) in the
respective filters. The results of
the fits with a Schechter function are given in Table 1. For the $r$ filter,
we also give the results of the fits when absolute magnitudes are computed 
from the corrected isophotal magnitudes. As can be seen, the determination of
the Schechter parameters is rather robust against the catalogs used. Differences
are always within $1\sigma$. In the following, we choose to adopt the Schechter
parameterization obtained from the catalogs limited in apparent magnitudes
to m$_r=21.5$, m$_g=21.8$, and m$_i=21.1$, which represents a fair compromise between
number of galaxies and correctness of background counts estimate. \\

\begin{table}
\caption{Full sample: LF best fit Schechter parameters}
\begin{tabular}{lllll}
\noalign{\medskip\hrule\medskip}

catalog & filter & $M^*$ & $\alpha$ & $\chi^{2}_{red}$/d.o.f.\\
limiting & & & &\\
magnitude & & & &\\

\noalign{\medskip\hrule\medskip}

22.7         & $r$ & $-21.39\pm0.10$ & $-0.87^{+0.10}_{-0.05}$ & 2.56/10 \\
21.5         & $r$ & $-21.36\pm0.10$ & $-0.84\pm0.08$ & 2.48/10 \\
21.0         & $r$ & $-21.32\pm0.10$ & $-0.82^{+0.10}_{-0.05}$ & 2.99/9 \\
\noalign{\bigskip}
22.5         & $g$ & $-21.02\pm0.10$ & $-0.87^{+0.04}_{-0.02}$ & 2.23/10 \\
21.8         & $g$ & $-20.97\pm0.10$ & $-0.82^{+0.05}_{-0.10}$ & 1.84/10 \\
21.3         & $g$ & $-20.99\pm0.10$ & $-0.83^{+0.08}_{-0.12}$ & 1.37/10\\
\noalign{\bigskip}
22.4         & $i$ & $-21.67^{+0.05}_{-0.10}$ & $-0.87^{+0.10}_{-0.05}$ & 2.33/10\\
21.1         & $i$ & $-21.62^{+0.05}_{-0.10}$ & $-0.83^{+0.08}_{-0.04}$ & 1.84/10\\
20.6         & $i$ & $-21.59\pm0.10$ & $-0.80\pm0.10$ & 1.75/10 \\
\noalign{\bigskip}
22.5         & $r^a$ & $-22.21_{-0.15}^{+0.10}$ & $-0.97\pm0.05$ & 2.16/10 \\
21.5         & $r^a$ & $-22.19^{+0.10}_{-0.15}$ & $-0.96_{-0.05}^{+0.07}$ & 2.32/9\\
21.0         & $r^a$ & $-22.16\pm0.15$ & $-0.95\pm0.07$ & 2.72/9 \\

\noalign{\medskip\hrule\medskip}
\end{tabular}

$^a$ isophotal magnitudes
\end{table}

Figure 6 (top to bottom) shows the composite luminosity
functions in the $i$, $r$ and $g$ bands (metric aperture 
magnitudes) obtained from the catalogs limited respectively to 21.1, 21.5 and
21.8 mag, together
with the 68 and 90\% confidence levels for the $M^*$, and $\alpha$ parameters
resulting from the fit of a Schechter function to the binned data. 
The normalization is arbitrary. It must be noticed that, although the
Schechter function is a fair representation of the composite LF in all
three bands, the quality of the fits is rather poor, several points
lying more than $1\sigma$ from the best fit value.\\

\begin{figure*}
\psfig{figure=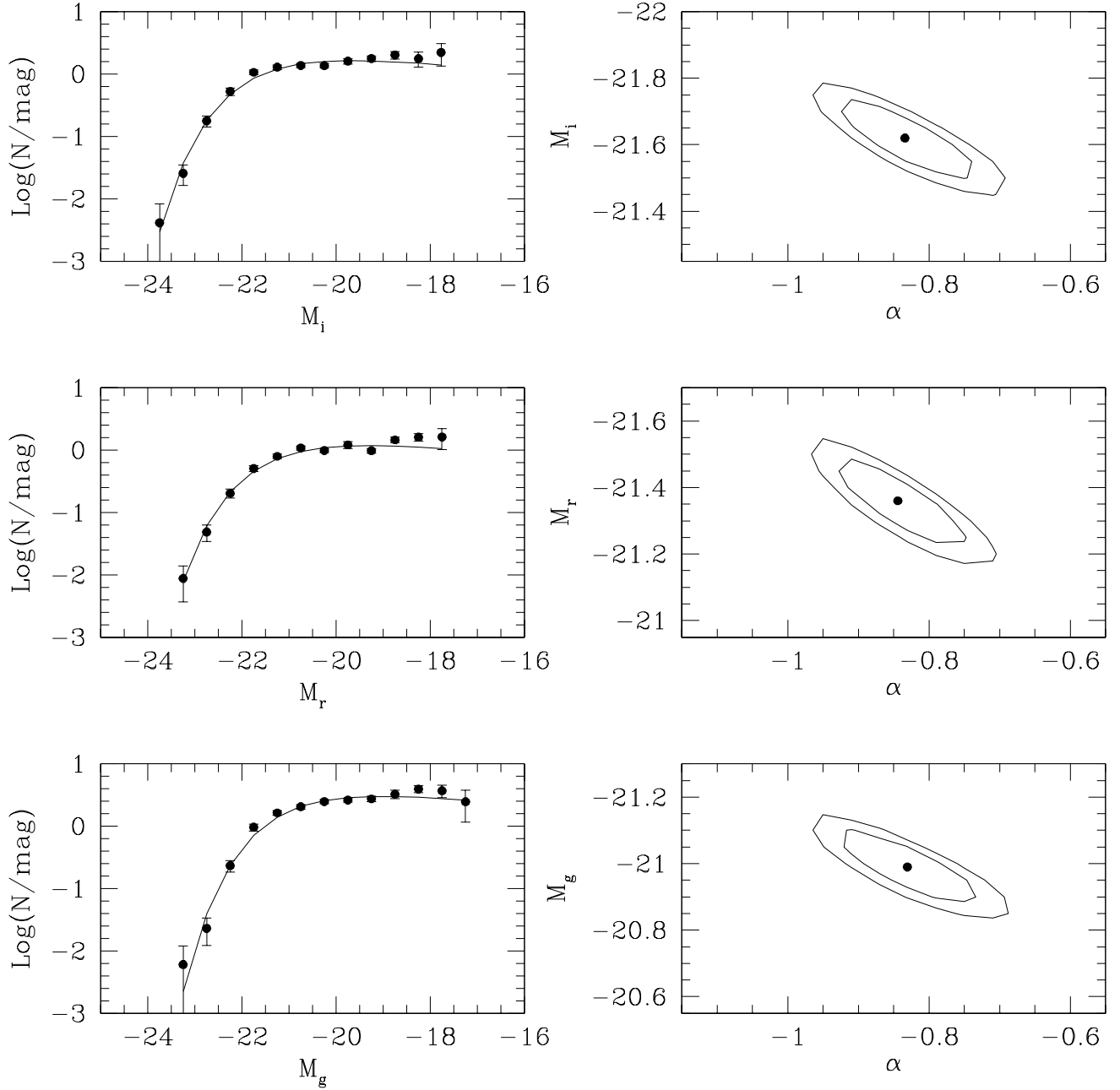,width=18cm,bbllx=20pt,bblly=160pt,bburx=570pt,bbury=700pt}
\caption{Composite cluster galaxy LF in the $i$, $r$ and $g$ bands (left panels)
and 90\% and 68\% confidence contour levels of the best fitting Schechter function
parameters. Absolute metric aperture magnitudes have been used. The galaxy catalogs 
for which the LFs have been derived were limited
to m$_i=21.1$, m$_r=21.5$, and m$_g=21.8$.}
\end{figure*}

To search
for differences in the LF depending on cluster properties, we have subdivided our data in various ways: galaxies in clusters at $z<0.15$ 
and at $z>0.15$,
galaxies in rich ($R\ge2$) and poor (R$\le1$) clusters (where all
EMSS clusters fall in this latter category), galaxies in X-ray selected (EMSS)
and in optically selected (Abell) clusters,
galaxies in early type (Bautz-Morgan types I and I-II) and
later type clusters, and galaxies in {\it dense} and {\it loose}
clusters. The latter subdivision has been obtained
by assuming that all clusters are described by a King (1962) profile

$$S(r) = {S_0\over{1+{(r/r_c)^2}}}$$

\noindent
where we set $r_c=250$ kpc. We then computed $S_0$
by integrating the radially symmetric
profile over each cluster field of view and by equating it to number of
galaxies
brighter than $M_r=-20.1$, roughly corresponding to the magnitude limit
utilized by
Dressler (1980) to study the morphology--density relation. Clusters have 
then been divided into {\it loose}
($S_0<S_0^{median}$) 
and {\it dense} ($S_0>S_0^{median}$) subsamples, where $S_0^{median}=74$
galaxies Mpc$^{-2}$. The average $S_0$
values of the two groups differ by a factor 2. The composite LFs for the subsamples have
been constructed in the same way as for the total sample.\\

We cannot search for the behavior of the LFs depending on cluster 
velocity dispersion, as this parameter is missing for most
of our clusters.\\

Table 2 shows the results obtained by fitting a
Schechter function to the data subdivided into subsamples.\\

\begin{table}
\caption{Subsamples: $r$ LF best fit Schechter parameters}
\begin{tabular}{llllll}
\noalign{\medskip\hrule\medskip}

subsample & $M^*$ & $\alpha$ & $M_{limit}$ & $\chi^{2}_{red}$/d.o.f. \\
\noalign{\medskip\hrule\medskip}

poor clusters & $-21.21\pm0.10$ & $-0.80\pm0.10$ & $-18.5$ & 2.28/7 \\
rich clusters & $-21.20^{+0.20}_{-0.10}$ & $-0.57\pm0.10$ & $-18.5$ & 1.32/7 \\
\noalign{\bigskip}
{\it loose} clusters & $-21.50\pm0.20$ & $-1.06^{+0.15}_{-0.10}$ & $-18.0$ & 1.80/8\\
{\it dense} clusters & $-21.19\pm0.10$ & $-0.59\pm0.10$ & $-18.5$ & 1.63/8 \\
\noalign{\bigskip}
Abell clusters & $-21.37\pm0.10$ & $-0.81_{-0.10}^{+0.05}$ & $-18.0$ & 2.83/8 \\
EMSS clusters  & $-21.11\pm0.20$ & $-0.82\pm0.20$ & $-18.0$ & 1.43/8 \\
\noalign{\bigskip}
BM I, I-II & $-21.27_{-0.10}^{+0.20}$ & $-0.82_{-0.15}^{+0.10}$ & $-18.0$ & 2.83/8 \\
BM II, II-III, III & $-21.44^{0.10}_{-0.20}$ & $-0.79_{-0.05}^{+0.15}$ & $-18.0$ & 1.73/8 \\
\noalign{\bigskip}
$z>0.15$ & $-21.20\pm0.20$ & $-0.57_{-0.10}^{+0.20}$ & $-19.0$ & 1.52/6 \\
$z<0.15$ & $-20.98^{+0.20}_{-0.10}$ & $-0.58\pm0.15$ & $-19.0$ & 1.75/5 \\
\noalign{\medskip\hrule\medskip}

\end{tabular}
\end{table}

\section{Discussion}
Let us first consider the LFs of cluster galaxies in the three bands.
As can be seen from Figure 6 (see also Table 1), the $g$, $r$ and $i$ LFs substantially have the same
shape. The magnitude shift of $M^*$ corresponds to the mean color
difference of early type galaxies in clusters.
The best fitting Schechter function shows a tendency to underestimate the
number of galaxies at luminosities around $M_r=-21$ and fainter than
$M_r=-19$, while it overestimates the number of galaxies with
intermediate luminosities. Quite some time ago, Oemler (1974) found slight
differences between
the LFs of spiral rich and spiral poor clusters. In a sample of 8 clusters,
Oegerle \& Hoessel (1989) found that the faint end of the LFs
varied between $<-1$ and $-1.25$, while the dispersion of $M^*$ was
only 0.24 mag. Previously, Lugger (1986) had concluded that the LFs of the
9 clusters she studied formed a rather uniform sample. At the same time,
Sandage, Binggeli \& Tammann (1985) decomposed the LF of the Virgo galaxies
into several morphological components fit by different functions, and
recently Biviano et al. (1996) showed that the Coma cluster LF is
better fit by a Gaussian and a Schechter function, dominating respectively
at the bright and faint end. If the general rule is that different cluster
LFs are fit by different Schechter functions or by a combination of Schechter
and other functions, it is to be expected that a composite LF would not
be nicely fit by a single function. As also Gaidos (1997) noticed in the
composite LF he derived, the Schechter function remains a fair representation
of the data, but the improved statistics with respect to earlier works
shows that the underlying hypothesis of the universality of the cluster LF
should probably be abandoned. The cluster morphological mix and the
morphology--density relation (Dressler 1980) should give rise to
LFs with different shape when subdividing a sample between galaxies in {\it dense}
and {\it loose} cluster environments, as Lugger (1989) found when
constructing the LFs of the inner and outer cluster regions.\\

\begin{figure}
\psfig{figure=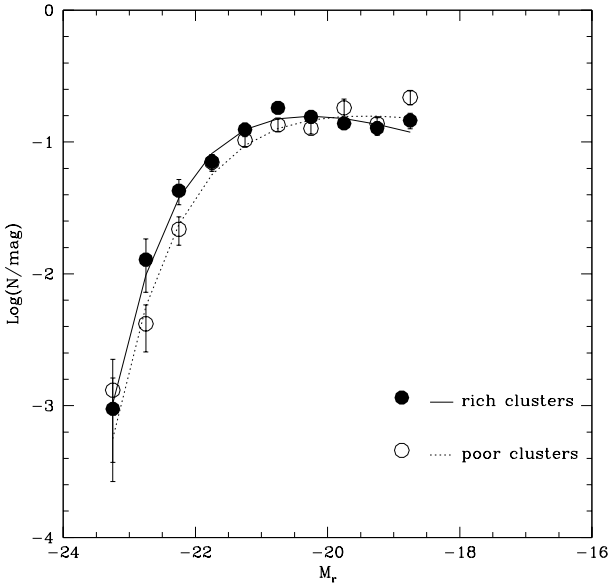,width=8cm,bbllx=20pt,bblly=160pt,bburx=570pt,bbury=700pt}
\caption{Composite cluster galaxy LF in the $r$ band for the rich (Abell richness
class $\geq2$ (data points: filled dots; best fit Schechter function: continuous line) and poor clusters (data points: open circles; best fit Schechter function:
dashed line). Each LF has been
normalized to the sum of the number of galaxies in each bin.}
\end{figure}

\begin{figure*}
\psfig{figure=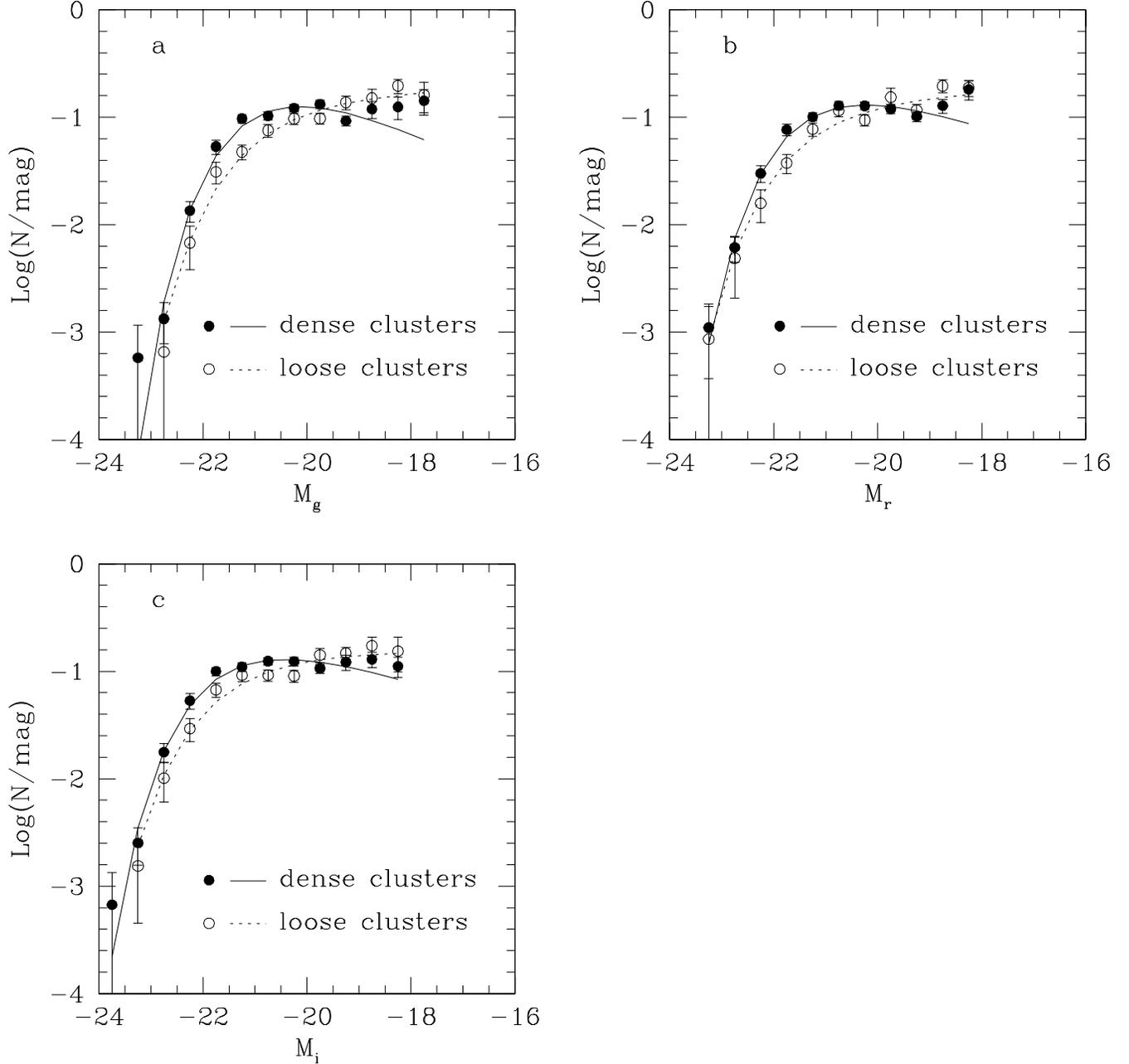,width=18cm,bbllx=20pt,bblly=160pt,bburx=570pt,bbury=700pt}
\caption{Composite cluster galaxy LF in the $g$ (panel $a$), $r$ (panel $b$) and
$i$ (panel $c$) bands for the $dense$ (filled dots; best fit Schechter function:
continuous line) and $loose$
clusters (data points: open circles; best fit Schechter function: dashed line). 
See the text for the definition of $loose$
and $dense$ clusters. Each LF has been normalized to the sum the number of
galaxies in each bin. }
\end{figure*}

Inspection of Table 2 shows that the poor and rich cluster LFs, in this case limited
to $M_r=-18.5$, the magnitude of the shallower subsample, differ in $\alpha$ at more than $1\sigma$, while $M^*$ does not change. The difference, though (see Figure 7), seems to be due more to a higher ratio
of $M^*$ galaxies to fainter ones in rich clusters with respect to poor ones, rather
than to a steepening of the poor cluster LF. This same type of difference 
is enhanced when considering the LFs of cluster galaxies in
the {\it dense} and {\it loose} subsamples. Figure 8
shows the LFs of galaxies in loose and dense environments in the three
bands. The Schechter fits are quite poor in some cases, especially
for the $g$ and $r$ band LFs of the {\it dense} subsample, and for the
$r$ band LF of the {\it loose} subsample.
Galaxies in loose environments
show similar LFs (with the expected color shift) in all three bands.
In order to quantify the difference in the LFs of {\it dense} and {\it loose}
cluster galaxies, we can consider the probability of finding a galaxy
brighter than $-20$ mag ($-19.5$ in the $g$ band). This probability
is always $>30\%$ higher in {\it dense} clusters. Thus, the density of
the environment is a factor which correlates with the galaxy luminosity
over a wide range of wavelengths. Not only are cDs found in
clusters and nowhere else, but bright galaxies in general are more
likely to be found in dense cluster environments.
Qualitatively, we can assume that, at some stage of cluster formation, 
the dense environment
stimulates merging or accretion phenomena and the formation of more luminous and,
perhaps, more massive galaxies. As already discussed by Lugger (1989), it is
not immediate to discriminate among the several phenomena known to occur
in clusters (merging, tidal stripping, morphological composition, infall,
subclustering), and which could be related to the density parameter. We
believe that real progress could be made by investigating
galaxy samples in well
controlled density environments with photometry extending into the near IR,
so that the mass distribution can be studied as well.\\

We find no evidence of an influence of the dynamical state of clusters on their galaxy
LF: early and late B-M types show the same LF, as do Abell and X-ray selected clusters.
We are thus led to believe that the cluster evolutionary stage or the selection
criterion are factors with little or no impact on the galaxy luminosity, or better,
they are not primary factors in determining the luminosity distribution of the
constituent galaxies, at least of the giant population in their cores.\\

The cluster galaxy LFs we obtained in the $r$ band can be directly
compared with the one obtained by Gaidos (1997) and the determinations of the
LFs measured in the field (apart from a normalization factor).
For this and the following comparisons we use the corrected isophotal magnitude instead of the metric aperture magnitude (see Table 1). Furthermore, we
assume $m_r=m_{R(KC)}+0.33$. Our 
{\it loose} cluster galaxy LF matches very well Gaidos' LF (the fit with a
Schechter function is strikingly similar, see also Figure 9). Of course, the {\it dense}
cluster galaxy LF differs from Gaidos', since it shows a flatter faint end
slope and an excess of bright galaxies. While our results confirm Gaidos'
value of $M^*$ in clusters, they also show that this value, or better, the LF
shape, is dependent
on galaxy density.\\

\begin{figure}
\psfig{figure=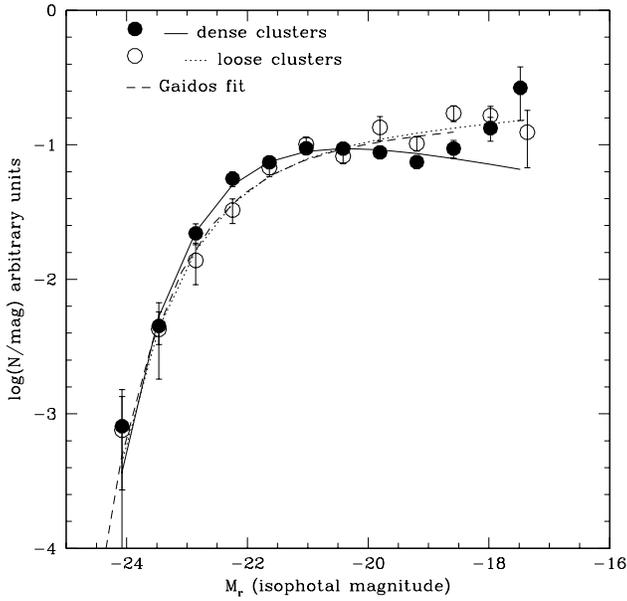,width=8cm,bbllx=35pt,bblly=160pt,bburx=570pt,bbury=700pt}
\caption{The composite LF of $loose$ and $dense$ clusters compared with the one obtained by Gaidos (1997) (long dashed line). Note that in this
case we used corrected isophotal magnitudes and not aperture magnitudes.}
\end{figure}

\begin{figure}
\psfig{figure=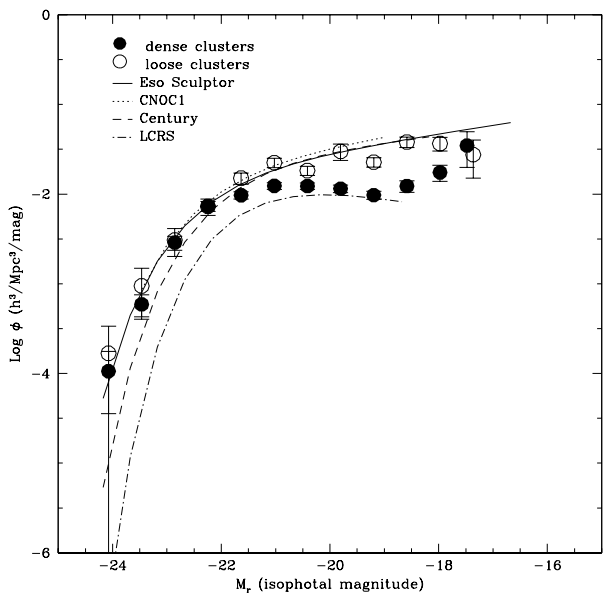,width=8cm,bbllx=35pt,bblly=160pt,bburx=570pt,bbury=700pt}
\caption{The composite $loose$ and $dense$ cluster galaxy LFs we obtained compared
with recent determinations of the galaxy LF in the field in the $R$ band. The
cluster galaxy LFs have been normalized at $M^*$ (corrected isophotal magnitudes)
to the average of the $\phi^*$ values given for the four field LFs.}   
\end{figure}

In the field, where galaxy density is lower, there are four recent
LF determinations in the $R$ band: from the LCRS (Lin et al. 1996), the CNOC1 redshift
survey (Lin et al. 1997),
the Century Survey (Geller et al. 1997) and the ESO-Sculptor redshift survey 
(de Lapparent et al. 1997). In Figure 10 we plot the four field LFs together 
with our cluster LFs data points normalized at $M^*$ to the average of the $\phi^*$ values given for the
four field LFs. The {\it loose} cluster galaxy data points are rather well described
by the ESO Sculptor and the CNOC1 LFs, but not so well by the Century LF and even
less well by the LCRS LF, which show a lack of bright galaxies. The {\it dense} cluster galaxy LF is dissimilar from any
field galaxy LF, as it is to be expected if the LF is density dependent. We believe that
progress on this issue could probably only be obtained by computing the LFs in
different density regimes with data showing the same type of selection biases:
in this sense the CNOC project seems to be the most promising.\\

\section{Conclusions}
We measured a composite cluster galaxy LF in three bands extending over 
more than 5 magnitudes.
We found that the LF of cluster galaxies is dependent on the a measure of
the density of the environment.
Bright galaxies ($M_r<-20$) have a higher probability of being found in
dense clusters. Some recent determinations of the field LF are rather similar
to the LF we measure in the less dense clusters.\\

\begin{acknowledgements}
We wish to thank the referee of this paper for the suggestions which
prompted us to improve the presentation of the results and of their discussion.
\end{acknowledgements}

\end{document}